\documentclass{elsarticle} % [review]

\usepackage[T1]{fontenc}
\usepackage[utf8]{inputenc}
\usepackage{subfig}
\usepackage{hyperref}
\usepackage{tikz}
\usepackage{amsmath, amsthm, amssymb}
\usepackage{amsfonts}
\usepackage{mathtools}

\title{Search by triplet: An efficient local track reconstruction algorithm for parallel architectures}

\author[1,2]{Daniel Hugo Cámpora Pérez\corref{cor1}}
\ead{dcampora@cern.ch}

\author[3]{Niko Neufeld}
\author[4,5]{Agustín Riscos Núñez}
\cortext[cor1]{Corresponding author\\\textit{Email address:} \href{mailto:dcampora@cern.ch}{dcampora@cern.ch}}
\address[1]{Maastricht University, Maastricht, Netherlands}
\address[2]{Nikhef National Institute for Subatomic Physics, Amsterdam, Netherlands}
\address[3]{CERN, CH-1211 Geneva 23, Geneva, Switzerland}
\address[4]{Smart Computer Systems Research and Engineering Lab (SCORE), Research Institute of Informatics Engineering (I3US), Universidad de Sevilla, Spain.}
\address[5]{Research Group on Natural Computing, Universidad de Sevilla, Spain}

\graphicspath{{figures/}}

\usetikzlibrary{backgrounds,fit}
\usetikzlibrary{shapes.geometric, arrows}
\usetikzlibrary{decorations.pathmorphing}
\usetikzlibrary{positioning}

\tikzstyle{arrow} = [thick,->,>=stealth]

\begin{document}

\begin{abstract}

Millions of particles are collided every second at the LHCb detector placed inside the Large Hadron Collider at CERN. The particles produced as a result of these collisions pass through various detecting devices which will produce a combined raw data rate of up to 40 Tbps by 2021. These data will be fed through a data acquisition system which reconstructs individual particles and filters the collision events in real time. This process will occur in a heterogeneous farm employing exclusively off-the-shelf CPU and GPU hardware, in a two stage process known as High Level Trigger.

The reconstruction of charged particle trajectories in physics detectors, also referred to as track reconstruction or tracking, determines the position, charge and momentum of particles as they pass through detectors. The Vertex Locator subdetector (VELO) is the closest such detector to the beamline, placed outside of the region where the LHCb magnet produces a sizable magnetic field. It is used to reconstruct straight particle trajectories which serve as seeds for reconstruction of other subdetectors and to locate collision vertices. The VELO subdetector will detect up to $10^9$ particles every second, which need to be reconstructed in real time in the High Level Trigger.

We present Search by triplet, an efficient track reconstruction algorithm. Our algorithm is designed to run efficiently across parallel architectures. We extend on previous work and explain the algorithm evolution since its inception. We show the scaling of our algorithm under various situations, and analyze its amortized time in terms of complexity for each of its constituent parts and profile its performance. Our algorithm is the current state-of-the-art in VELO track reconstruction on SIMT architectures, and we qualify its improvements over previous results.

\end{abstract}


\maketitle

\section{Introduction}
\label{sec:introduction}

% The LHCb detector is a large physics detector situated at the Large Hadron Collider at CERN. Millions of particle collisions occur every second inside the detector, which produce particles that traverse the detector. The velocity, energy and momentum of produced particles can be detected through reconstructing the signals left in the LHCb subdetectors, which serve to collect statistical information and ultimately infer the types of processes that occur at a fundamental level.

% Data incoming from the LHCb detector are piped through a data acquisition system that performs a data filtering in real-time. The LHCb detector and its data acquisition system are being upgraded in preparations for the restart of data taking scheduled for 2021. 

The LHCb detector is a large physics detector situated at the Large Hadron Collider at CERN~\cite{LHCbCollaboration2015}. The detector is being upgraded for the restart of data taking scheduled for 2021~\cite{LHCbTriggerOnlineTDR:2014}. The full collision data rate of 40~Tbps will be piped through a data acquisition system that will perform a data filtering in real-time, prior to storing data in long-term storage for posterior analysis. The filtering will occur in two stages: the first stage or \emph{High Level Trigger 1} (HLT1) will reduce the data rate according to particle kinematics by a factor of $40\times$ in a computing farm composed of 170~servers equipped with GPUs~\cite{GPUTDR:2020}. The second filter stage or \emph{High Level Trigger 2} (HLT2) will perform a full event\footnote{An \emph{event} corresponds to a single crossing of the Large Hadron Collider proton beams.} reconstruction and reduce data by an additional factor~$20\times$ in a computing farm composed of thousands of servers~\cite{LHCbComputingTDR:2018}. The introduction of a heterogeneous computing infrastructure in LHCb is motivating the creation of parallel algorithms that are portable and efficient across architectures.

% Instead of "through a detector" - "under the potential presence of a magnetic field"

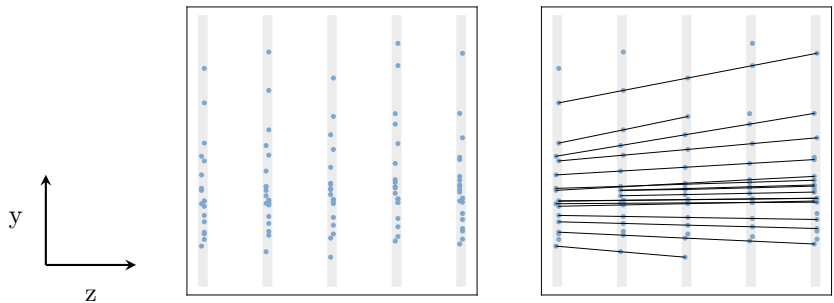
\begin{figure}[!hbt]
\begin{tikzpicture}[x=4mm, y=4mm]
\draw [arrow] (0,0) -- (0,3);
\draw [arrow] (0,0) -- (3,0);
\node at (1.5,-1) {z};
\node at (-1,1.5) {y};
\end{tikzpicture}
\quad
\includegraphics[width=0.35\textwidth]{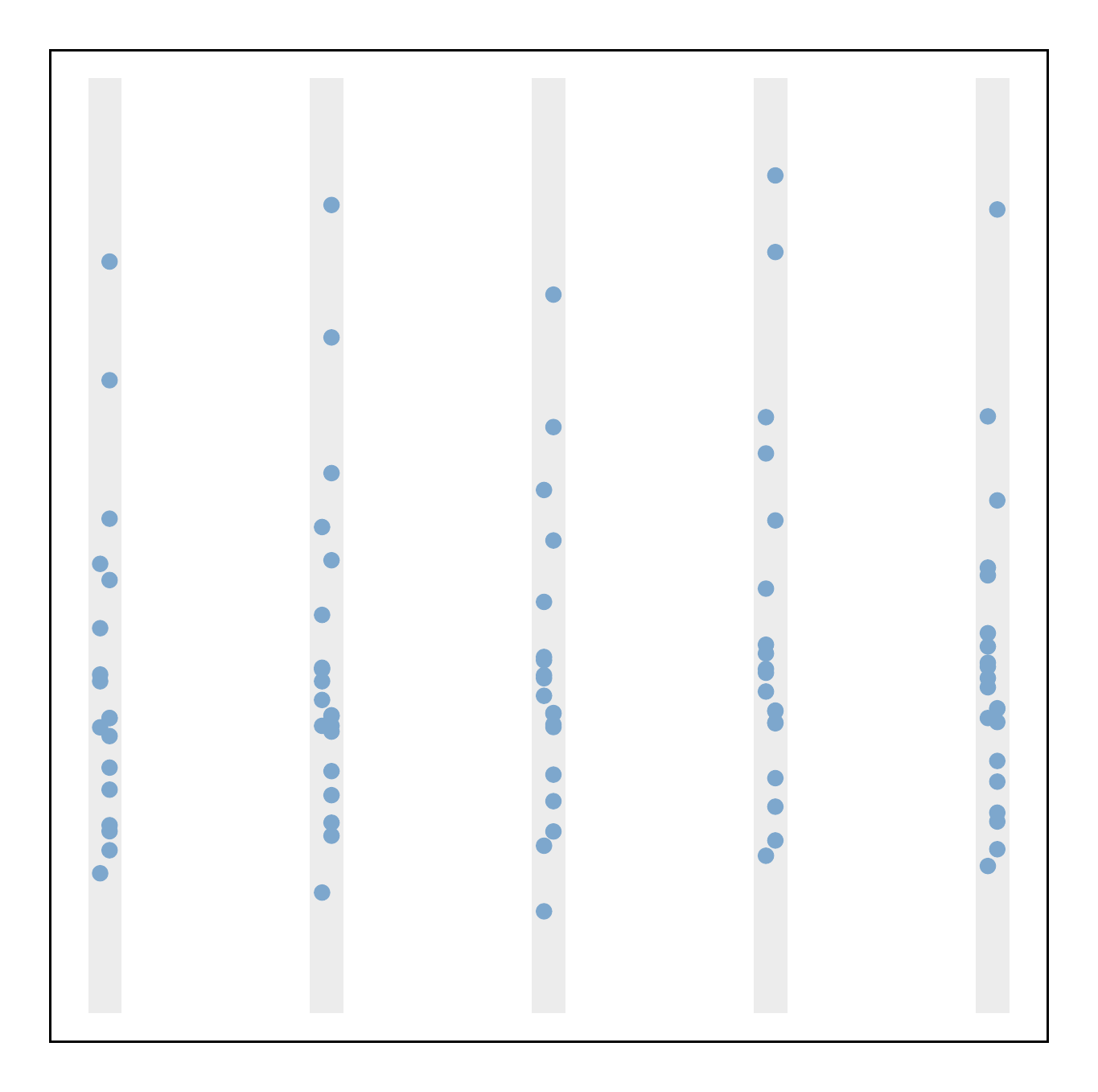}
\quad
\includegraphics[width=0.35\textwidth]{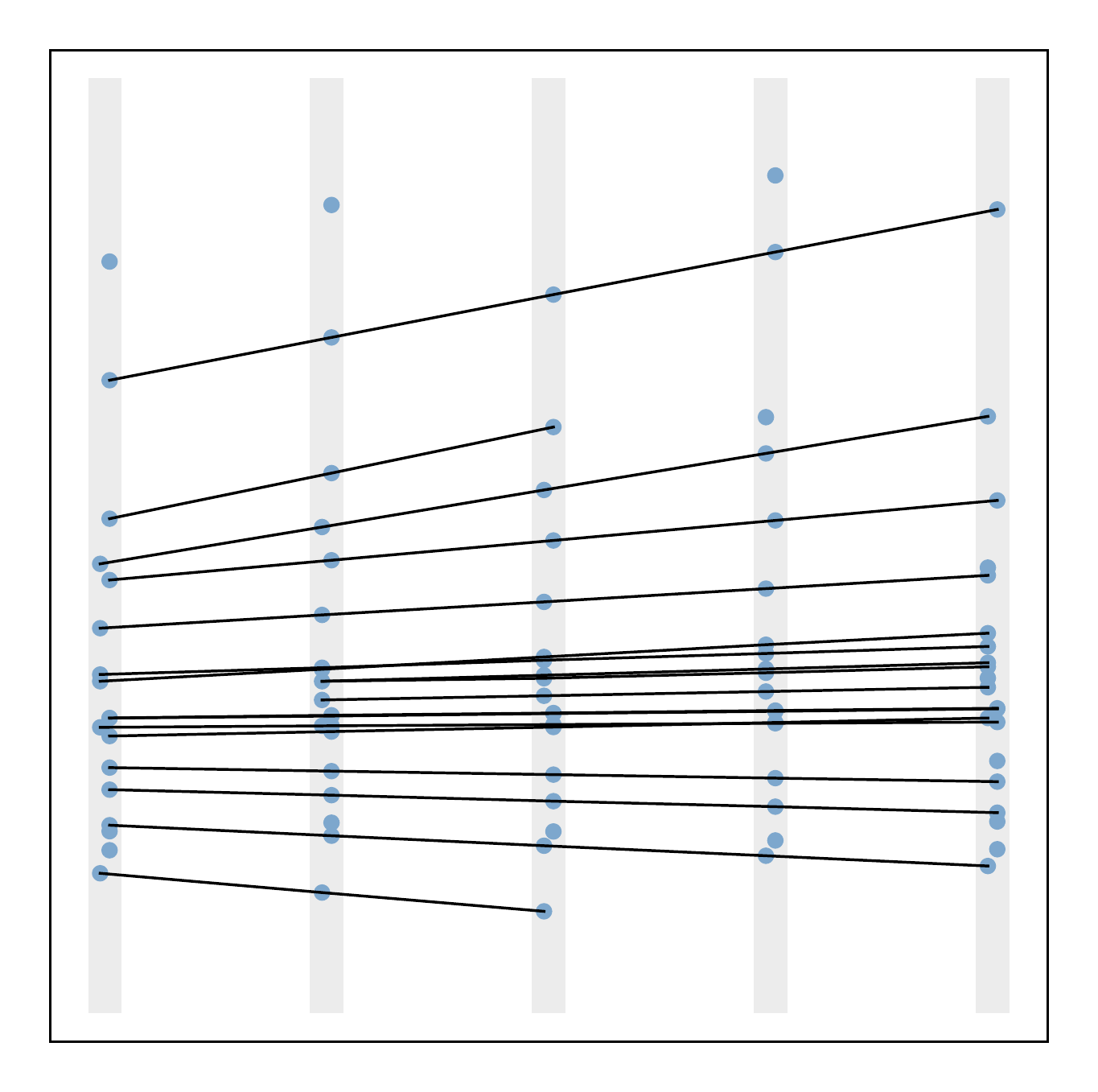}
\caption{(Left) A minimal track reconstruction instance projected in 2D, consisting in a set of hits with position information. (Right) The actual particle trajectories sought when doing track reconstruction.}
\label{fig:tracking-example}
\end{figure}

Track reconstruction or \emph{tracking} is a pattern recognition problem consisting in finding particle trajectories from measurements (\emph{hits}) in detectors along their path. The problem is equivalent to finding a partition of disjoint sets of measurements that are compatible with the laws of motion of particles as they traverse a detector, accounting for the fact that some measurements may be noise, and considering the presence of sizable magnetic fields which curves the trajectories of charged particles depending on their momentum. Track reconstruction yields momentum and trajectory information of reconstructed particles, which play an essential role in trigger systems of physics experiments. Figure~\ref{fig:tracking-example} exemplifies the track reconstruction problem.

% Tracking example

\begin{figure}[!hbt]
  \centering
  \includegraphics[width=\textwidth]{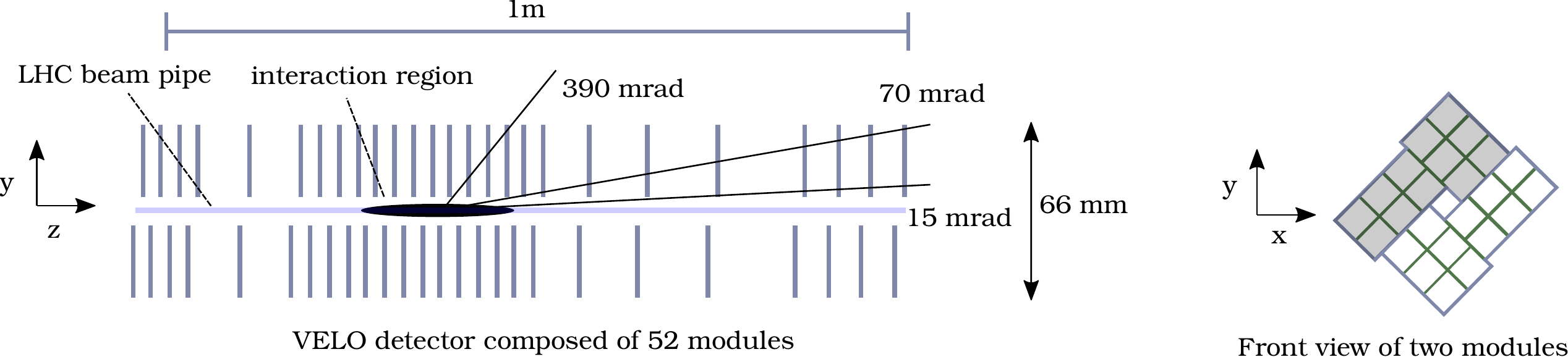}
  \caption{A schematic of the upgraded Velo detector. (Left) Section in the XZ plane, with detector modules laying in two sides. (Right) Front view of each module in the XY plane, with subdivisions indicating detector chips. Each detector chip has a resolution of $256 \times 256$~pixels.}
  \label{fig:velo-subdetector}
\end{figure}

The Vertex Locator (VELO) is a tracking detector of LHCb consisting of 52~planes of silicon pixel chips surrounding the LHC interaction point and beamline, shown in Figure~\ref{fig:velo-subdetector}. As particles pass through the detection planes, they leave detectable measurements in the form of pixel clusters. VELO track reconstruction constitutes the first reconstructed subdetector of LHCb, and tracks found in the VELO are used to locate the originating collision vertices as well as serve as seeds for subsequent track reconstruction. Therefore reconstructing the VELO is of vital importance towards the correct functioning of LHCb.

% Image of the VELO detector

Track reconstruction validation is typically performed with Monte Carlo (MC) simulated samples, where reconstructed tracks should match MC particles, which establish the \emph{ground truth}. The matching of tracks with particles is done on a hit by hit basis. The physics performance of found tracks can be evaluated according to five indicators~\cite{Schiller2011}:

\begin{itemize}
\item The track \emph{reconstruction efficiency} can be determined by the ratio between the reconstructed tracks of reconstructible particles, over all the reconstructible particles\footnote{A particle is said to be \emph{reconstructible} in a tracking detector when it fires enough measurements over a detector-specific threshold.}:

\begin{equation}
    \dfrac{N_{\text{reconstructed and reconstructible}}}{N_{\text{reconstructible}}}
\end{equation}

\item A \emph{fake track} (\emph{ghost track}) is created when a percentage of hits in a track are not from a real track. In LHCb, at least 70\% of the hits in a track must belong to the same MC particle to be associated in the validation process. The \emph{fake track fraction} is the ratio between the fake tracks and all the reconstructed tracks:

\begin{equation}
    \dfrac{N_{\text{fake tracks}}}{N_{\text{reconstructed tracks}}}
\end{equation}

\item The \emph{clone track fraction} refers to the fraction of tracks associated to the same MC particle as another reconstructed track:

\begin{equation}
    \dfrac{N_{\text{clone tracks}}}{N_{\text{reconstructed tracks}}}
\end{equation}

\item The \emph{hit purity} in a track refers to the fraction of track hits that belong to the same MC particle:

\begin{equation}
    \dfrac{N_{\text{track hits in MC particle hits}}}{N_{\text{track hits}}}
\end{equation}

\item Finally, the \emph{hit efficiency} yields the number of hits correctly found out of the MC particle hits in a track:

\begin{equation}
    \dfrac{N_{\text{track hits in MC particle hits}}}{N_{\text{MC particle hits}}}
\end{equation}
\end{itemize}

%  as evaluated by the performance indicators
% within the hardware constraints of the data acquisition system
The VELO detector is outside of the range of effect of the LHCb magnet, and therefore trajectories can be considered to be straight lines. Once the detector restarts operation a sustained throughput of $10^9$ particle trajectories at the VELO per second will have to be reconstructed in the trigger systems, while delivering good track reconstruction performance indicators. The VELO reconstruction is therefore a real-time software challenge whereby the design performance of the system must be met within the hardware constraints of the data acquisition system.

\section{Tracking techniques}

% various track reconstruction techniques have been documented in literature

Due to the interest in tracking by many particle physics experiments, there is a rich literature on track reconstruction techniques~\cite{Fruhwirth2000}. Local tracking methods find tracks iteratively, whereas global methods adapt an equivalent formulation of the problem, typically including all measurements, where solutions map to tracks.

The most common local tracking method consists in finding a \emph{track seed} and extending it to other detector planes in a process known as \emph{track following}. The track seed is usually formed by a segment of two or three hits, and the search starts in a region where the hit density is lower and thus the signal is easier to distinguish, which usually corresponds with the furthest distance to the expected interaction point. Track seeds are extrapolated (\emph{followed}) to detector regions closer to the interaction point by applying an extrapolation accounting for the presence of a field if necessary. A model of the track can be formed from the track hits, and this model can be employed to select among a list of candidate hits the best fitting one. This extrapolation process may account for missing hits in detector parts, according to the hit inefficiencies of the physical detector and \emph{dead} regions without sensitive detectors. Once a track is fully built, its constituent hits can be flagged so they are not revisited in further seed or following steps.

Historically a local track following method has been employed for the VELO reconstruction of the LHCb experiment runs, named \emph{Search by pair}~\cite{Callot2011}. It constructs seeds of pairs of hits initially in the furthest modules from the interaction region. Track seeds are followed to neighboring modules, allowing for for one hit to be missing in consecutive modules on any one side of the VELO subdetector. Tracks of four or more hits flag all of their hits, reducing the search time of further seeding and following steps. Tracks consisting of at least three hits are stored, in accordance to the reconstructibility condition of the VELO subdetector. However, the requirement to flag hits as used makes this technique not suitable for parallelization without modification. It is possible to find all triplet combinations in parallel dropping the flagging mechanism, which was done in the seeding phase of~\cite{Funke2014a}. However this parallelism comes at the cost of generating all possible triplet combinations, which is inefficient for densely populated detectors.

Other local tracking methods are \emph{track roads} and \emph{track elements}. \emph{Track roads} forms candidates with two hits situated in the extremes of the detector, and creates a path or \emph{road} between both hits by interpolation. In case the model of the track be curved, a third hit should be added. The width of the road determines the accepted error in the model, and it depends on the characteristics of the detector~\cite{Frohlich1976}. \emph{Track elements} performs reconstruction in two phases: (1)~seeds are made up from neighboring points, straight lines or parabolic lines. Each seed is converted into a \emph{master point} (a weighted average of the points) and a direction. (2) The seeds, instead of the original hits, are used to perform tracking. This method reduces the number of hits to consider in the tracking phase, at the cost of a loss in precision~\cite{Olsson1980}. Local methods may be used in conjunction with corrections such as the Kalman filter~\cite{Kalman1960}, fitting forming tracks and selecting hits~\cite{CamporaPerez2018a}.

% Removed at the end: , as in the ALICE TPC sectors
Hit multiplicity is often a concern in real-time reconstruction environments, where track reconstruction must be performed at a high throughput in order to keep up with the collision rate. Spatial reductions can be employed to reduce the search time of hits under consideration for local tracking methods. This envolves a data preparation step prior to the application of the tracking method. Common data structures used alongside local methods include KD-trees~\cite{Lopes2014}, binary search structures or more generically search windows. The specifics of the geometry of the detector yield in some cases a natural subdivision of the problem~\cite{Rohr2012}.

% Finding all compatible triplets can be parallelized dropping the flagging mechanism, like in the seeding phase of~\cite{Funke:2014}. However, this is inefficient for densely populated detectors. Local methods are commonly used in conjunction with an estimator like the Kalman filter~\cite{Kalman:1960} to fit forming tracks and select hits~\cite{Campora:2018}. Spatial reductions like KD-tree structures~\cite{Lopes:2014} or search windows help reduce the dimensionality of hits under consideration.

% Before: Alternatively, track reconstruction can be adapted to global formulations of the problem.
Alternatively, global methods can be applied to the track reconstruction problem. The \emph{Hough transform} method~\cite{Kalviainen1995} in its simplest form converts all hit points into a histogram representation in polar coordinates, where peaks are equivalent to compatible hits. Despite of the elegance of the Hough transform underlying principle, difficulties arise when dealing with high hit multiplicities, where binning and threshold of the histogram play an important role in avoiding excess of clones or fake tracks. Circular trajectories must be converted to lines prior to applying the Hough transform, which can be achieved with a \emph{conformal mapping} transformation~\cite{Yepes1996}.

The \emph{clustering method} consists in extrapolating hits onto a parameter space, according to the expected trajectory from a collision vertex. Once in the parameter space, hits pertaining to the same track appear close to each other. The tracking problem consists then in a clustering problem, that can be solved with any clustering method~\cite{Milligan1987}. The collision vertex must be chosen with sufficient precision, and the resulting cluster must be verified to match reasonable track parameters~\cite{Eichinger1980}.

The \emph{automata method}~\cite{Glazov1993} is a graph traversal method of a weighted directed graph representing the measurements. Each measurement is a vertex, and directed edges between measurements in neighboring detector elements are created according to the expected track parameters in the region of acceptance of the detector, in the direction from outer to inner-most layer. The graph is then traversed following a Depth-First Search, assigning weights to visited edges according to the current depth level. If vertices are visited multiple times, the highest weight prevails. The automata method has been successfully implemented in various trigger systems~\cite{Funke2014a, Rohr2017}.

% Note: The retina algorithm has not been cited nor explained.

% The Retina algorithm~\cite{Abba:2016} builds a heatmap for each hit to determine compatible tracks. The \emph{automata} technique~\cite{Kisel:1993} consists in creating a weighted graph representing the connectivity of every hit, and traversing it to find the best tracks.

% TODO: Cite 
%  Rohr:2017ydv - automata, current global method in ALICE
%  Seeding of Renato - histogramming method in LHCb
%  Patatrack
Both the algorithm and the target hardware architecture must be considered when devising an efficient track reconstruction solution. It has been discussed that global methods are amenable by design for parallel architectures~\cite{Fruhwirth2000, Eichinger1980}. Some of the high-throughput tracking algorithms used in current trigger systems are global methods~\cite{Rohr2017, Quagliani2017}, and have been implemented on modern CPU and GPU architectures. On the other hand, local methods have also been used for high-throughput solutions~\cite{ACTS2019}.

% Change this to argue that local methods are also promising for parallel architectures
% Talk about:
%  My previous publication, a local method that is efficient on gpu and cpu architectures
%  Mention Arthur's paper as a buildup on the foundation of the previous specifically for simd architectures
%  Argue that we are presenting an extension to our previous work for SPMD architectures, with architecture-specific SIMD optimizations
In our previous work~\cite{CamporaPerez2019}, we presented \emph{Search by triplet}, a novel local method for track reconstruction on parallel architectures, alongside a base framework to run it efficiently on GPU architectures. Our algorithm considers module pairs to increase the amount of parallel work in every step. We sort the data by $\varphi$ on every module pair stemming from the observation that hits of tracks have similar $\varphi$ values. Our algorithm then iteratively generates track seeds of triplets of hits, and extrapolates forming tracks to the next module pair. Fake and clone tracks are avoided by introducing a flagging mechanism with barriers. In a last step, track seeds that were never extended are checked again prior to storing them as tracks. We implemented our work both in the SIMT\footnote{\emph{SIMD} (Single Instruction Multiple Data) is a class of parallel processors in which each instruction operates over several data simultaneously in lockstep. \emph{SPMD} (Single Program Multiple Data) is a class of processors in which the work is distributed in tasks that operate over multiple data, not necessarily in lockstep. \emph{SIMT} (Single Instruction Multiple Thread) is an execution model employed by GPUs where every instruction is executed by a group of threads.} programming paradigm targeting GPUs, as well as in SPMD targeting CPUs. A variation of this algorithm with SIMD-specific optimizations was done in~\cite{Hennequin2019}.

In this paper, we build on the foundation of our previous algorithm and optimize it for SIMT and SIMD architectures. We improve the maintainability of our algorithm by providing a single codebase for any heterogeneous architecture in the Allen framework~\cite{GPUTDR:2020}. We also provide architecture-specific optimizations for routines that are employed often and that require hand-tuning. Through an iterative process of optimization, we increase the throughput of our algorithm by more than a factor~$3 \times$, while also improving its physics efficiency.

% Our algorithm meets the HLT design performance and constitutes the current state-of-the-art in VELO reconstruction.

\section{Search by triplet}
\label{sec:design}

Search by triplet is a local track following algorithm optimized to reconstruct the LHCb VELO detector that exploits the task parallelism inherent to the LHCb data taking regime and the data parallelism of track reconstruction. 30~million events are detected per second in LHCb, where each is independent of each other. Therefore, we assign different tasks to process each individual event. Within each event processing, track reconstruction exposes various levels of parallelism that we tackle in a data parallel fashion, either employing the SIMT paradigm or SIMD optimizations where relevant.

Search by triplet is now a single coherent codebase written in C++, with custom extensions that provide SIMT functionality. Depending on the target device specified at compile time, our codebase can potentially target any parallel architecture, notably modern CPUs and GPUs. Table~\ref{tab:allen-parallelism} depicts the levels of parallelism exploited in Allen algorithms such as Search by triplet, and the correspondance with each target.

\begin{table}[hbt!]
\begin{center}
\resizebox{\columnwidth}{!}{%
\begin{tabular}[!hbt]{|l|c|c|c|}
\hline
 & CPU target & GPU target & Type of parallelism \\\hline
Bunches of events & Threads & Streams & Task parallelism \\\hline
Inter-event & Sequential & Blocks & Task parallelism \\\hline
Intra-event & Sequential or vectorization & Threads & Data parallelism \\\hline
\end{tabular}
}
\caption{Degrees of parallelism for event reconstruction in Allen. CUDA terminology is used~\cite{Lindholm2008} for a generic GPU target, not necessarily NVIDIA.}
\label{tab:allen-parallelism}
\end{center}
\end{table}

Our algorithm consists in three sub-algorithms that are described in the following. For simplicity, we will refer to modules instead of module pairs. When discussing computational complexity, the notation employed is generalized to a detector with $m$ consecutive detector modules and an average number of hits in each module of $n$.

\subsection*{Sort by phi}

The VELO modules are positioned such that particle collisions tend to occur close to the origin of coordinates in the $XY$ plane. Hence, particles produced in these collisions that travel in a straight line are likely to have a constant phase in polar coordinates when projected on an $XY$ plane. Hits in each module are sorted by $\varphi$, calculated as the 2-argument arctangent in the $XY$ plane with respect to the origin of coordinates.

The range of values of the 2-argument arctangent is $[-\pi, +\pi]$. However, values close to $-\pi$ are conceptually very close to values close to $+\pi$, as they represent angles that are close. For this reason, we convert the range into a unsigned 16-bit integer type, mapping the 2-argument arctangent range to the unsigned 16-bit range $[0, 2^{16} - 1]$. This transformation enables us to use modulo arithmetic to perform comparisons, and it allows us to reduce memory pressure by using two bytes as opposed to four to store each $\varphi$.

Given that the number of maximum hits in a module is known, a constant amount of \emph{shared memory}\footnote{Shared memory is a GPU-specific optimization that uses the configurable L1-cache shared memory available in GPUs. For other targets, main memory is employed.} is employed as a means of low-latency temporary data buffer when $\varphi$ is calculated and sorted. The permutation produced is then employed to sort hit coordinates, yielding a structure of arrays sorted by phi for each module. A parallel insertion sort method has been implemented to calculate the permutation. The worst-case complexity of this algorithm is $O(m \cdot n^2)$.

An SIMD-specific optimization is also provided. The $\varphi$ of every hit is calculated using a vectorization library. The permutation is calculated using the quicksort implementation provided in the STL library. The STL sort implementation has a worst case of $O(n \cdot log(n))$, and therefore the SIMD specialization has a worst-case complexity of $O(m \cdot n \cdot log(n))$.

\subsection*{Track seeding and track following}

\begin{figure}[hbt!]
\centering
\includegraphics[width=\textwidth]{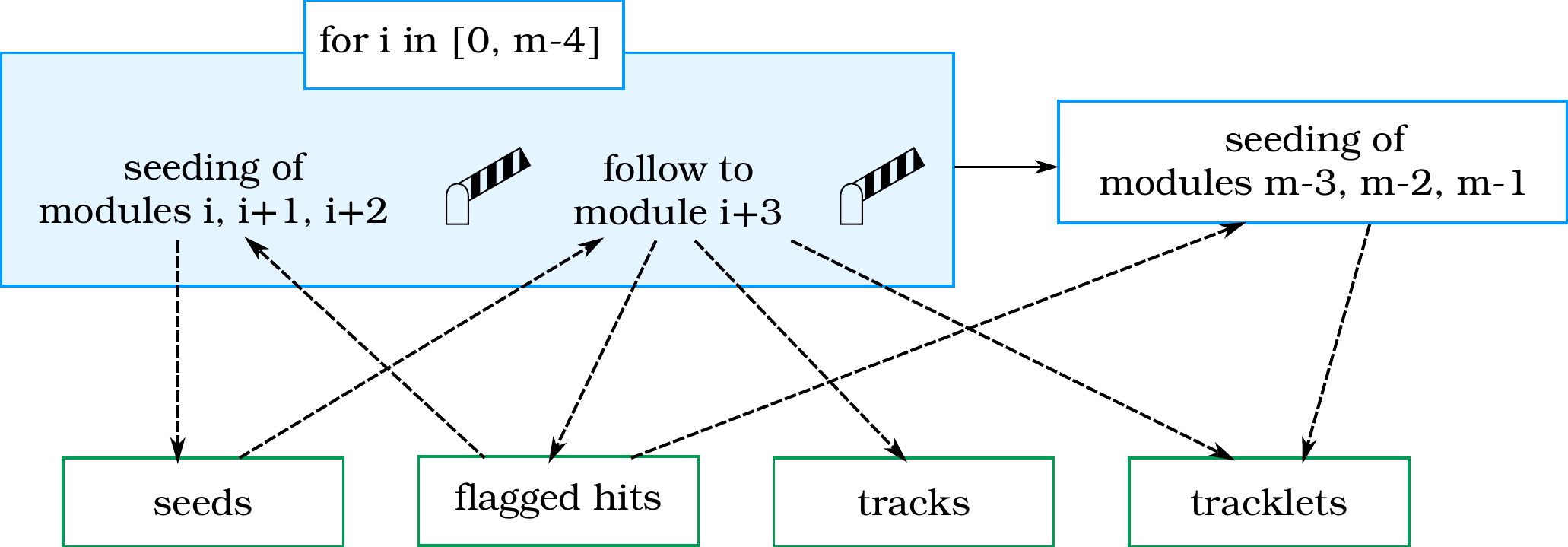}
\caption{(Top) Control and (bottom) data flow of the track seeding and track following steps. Only data containers relevant to the control flow are shown. SIMT barriers are represented with the icon \protect\includegraphics[width=14pt]{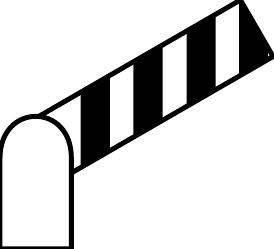}.}
\label{fig:control_flow}
\end{figure}

The local track reconstruction technique implemented in Search by triplet consists in an iterative application of two stages: track seeding and track following. Figure~\ref{fig:control_flow} shows an overview of the technique with the iterative control flow on the top, and the data containers involved in either data dependencies or output at the bottom. Even though the iteration follows an ascending order, the actual direction of traversal of the modules is configurable.

Track seeding finds triplets of hits in consecutive modules, one hit in each module. In order to avoid clone tracks, a flagging mechanism has been implemented, akin to the one found in \emph{Search by pair}~\cite{Callot2011}. None of the hits in the triplet must be flagged. This introduces the requirement of a barrier between each seeding and following stage due to the Read-After-Write and Write-After-Read dependencies: flagged hits are read during track seeding, and they are written during track following. The barrier is only present in SIMT code, and if an SIMD target is specified, it is removed at compilation time. The \emph{flagged hits} container is initialized to be empty prior to the execution of the first seeding iteration.

\begin{figure}[hbt!]
\centering
\includegraphics[width=0.4\textwidth]{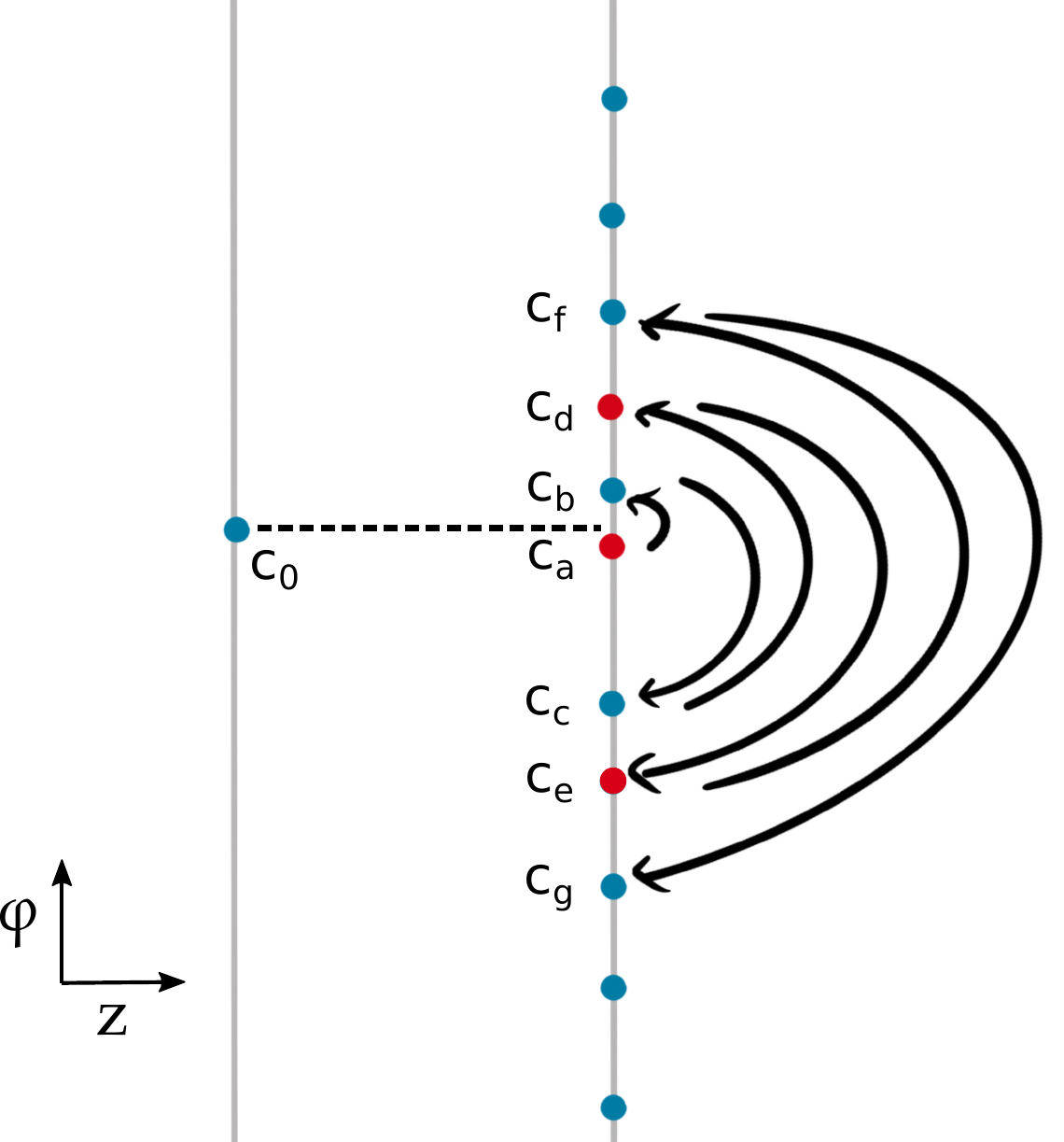}
\caption{\emph{Pendulum search} of the first four non-flagged hits closest to the $\varphi$ of hit $c_0$. Flagged hits are marked in red ($c_a$, $c_d$ and $c_e$).}
\label{fig:pendulum_search}
\end{figure}

The process of finding triplet seeds has received several optimizations. We consider three consecutive modules, and process all non-flagged hits in the middle module separately. For each hit, we seek the corresponding $\varphi$ position in the neighbouring module with a binary search. Then, we perform a \emph{pendulum search}, alternatively looking further down and up the hits in the module, until a threshold number of candidate hits that are not flagged is found. Figure~\ref{fig:pendulum_search} depicts this process.

\begin{figure}[hbt!]
\begin{center}
  \subfloat[][]{\includegraphics[width=0.42\textwidth]{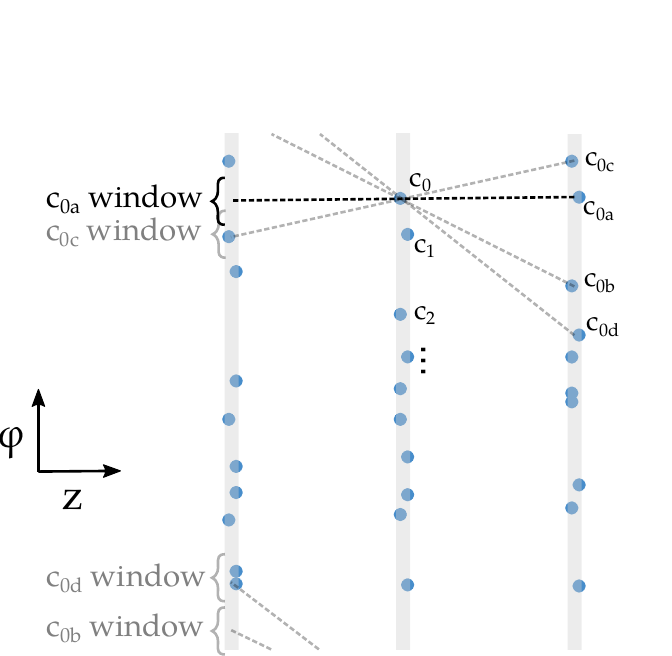}\label{fig:seeding}}\qquad\qquad
  \subfloat[][]{\includegraphics[width=0.42\textwidth]{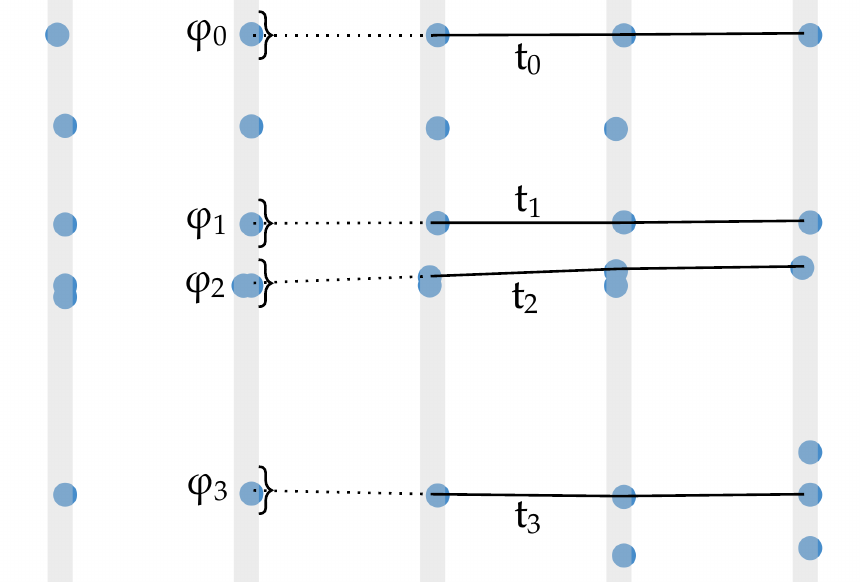}\label{fig:trace-a}}\\\vspace{4pt}
  \subfloat[][]{\includegraphics[width=0.42\textwidth]{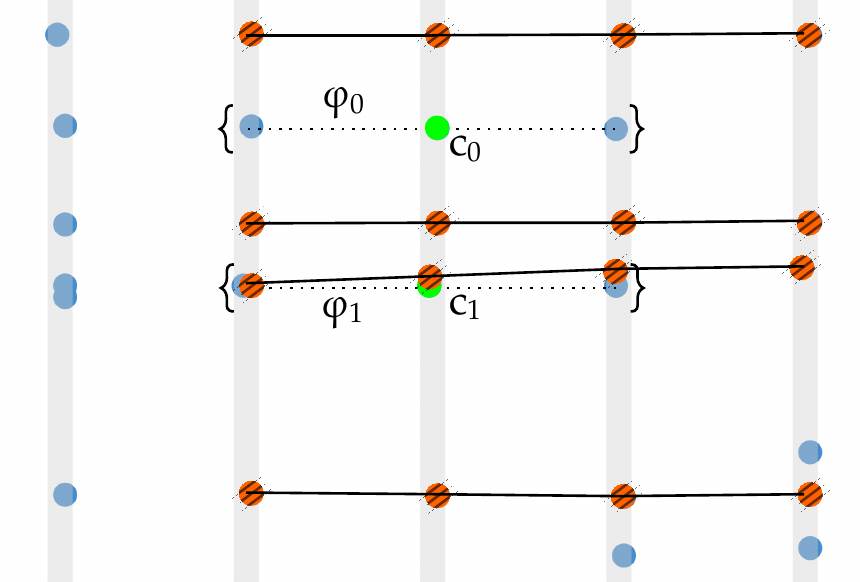}\label{fig:trace-b}}\qquad\qquad
  \subfloat[][]{\includegraphics[width=0.42\textwidth]{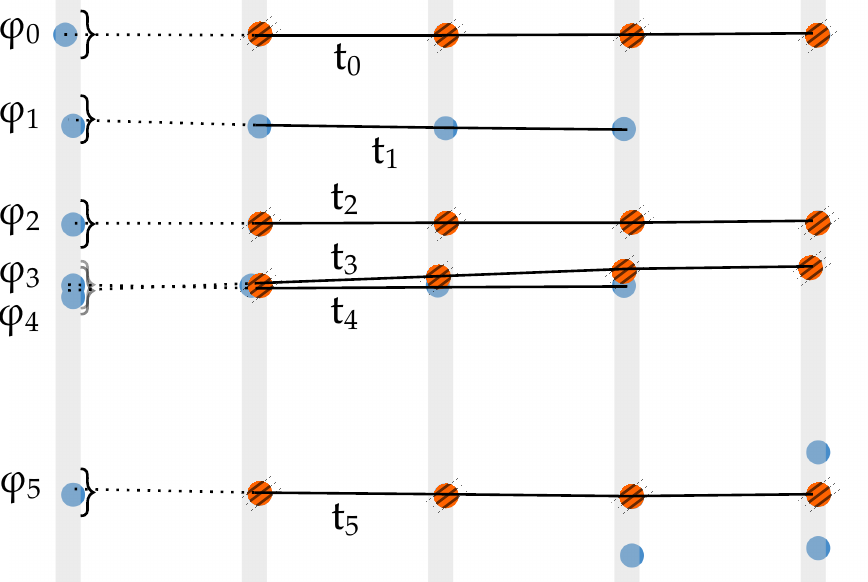}\label{fig:trace-c}}
\end{center}
\caption{Iterative seeding and following stages, where modules are considered from right to left. \protect\subref{fig:seeding} Seeding stage. For hit $c_0$, four hits $c_{0a}$, $c_{0b}$, $c_{0c}$ and $c_{0d}$ are considered on the neighbouring module on the right. Each of the resulting \emph{doublets} is extrapolated onto the neighbouring module on the left, where hits in the $\varphi$ window are considered. The $\varphi$ search wraps around. \protect\subref{fig:trace-a} Following stage. Forming tracks are extrapolated and hits are sought in a $\varphi$ window. \protect\subref{fig:trace-b} and \protect\subref{fig:trace-c} Subsequent seeding and following stages. Hits found in previous follow stages are marked as flagged and not further considered.}
\label{fig:trace}
\end{figure}

Once all candidate doublets are found, each doublet is extrapolated to the third module. A binary search is performed with the extrapolation $\varphi$, and all hits within a tolerance window are considered. Figure~\ref{fig:seeding} shows the process of extrapolation for hit $c_0$. Each triplet candidate is evaluated with a \emph{scatter} function between the extrapolated doublet position and the third hit under consideration: $dx^2 + dy^2$. For each middle module hit, at most one triplet is kept: the triplet minimizing the scatter function under a threshold.

Triplets are stored in the \emph{seeds} container and passed on to the following stage, where all forming tracks (in containers \emph{seeds} and \emph{tracks}) are considered. Track following consists in extrapolating forming tracks to the next module and attempting to find compatible hits under a threshold. Similarly to the doublet extrapolation of seeding, the last two hits of each forming track are extrapolated to the next module, and the best hit is evaluated with the scatter function. In order to account for the hit inefficiency of modules, it is allowed for tracks to miss a module, in which case they are kept for the next following iteration.

The best hits found for the followed tracks are flagged, as shown in Figure~\ref{fig:trace}. When a forming track misses two consecutive modules, or when processing the last iteration, forming tracks are either stored in the \emph{tracklets} or \emph{tracks} container, depending on whether the number of hits is three or more respectively.

The data parallelism of this process is exploited by separate threads in the SIMT model, whereby each thread processes a different starting hit in the seeding stage, or a different forming track in the following stage. An SIMD-specific optimization has been done for the seeding stage, whereby all doublets of each starting hit are processed in parallel with vector instructions.

The worst-case complexities of the two stages are as follows. Seeding is performed on all consecutive module triplets (bound by $m$), where all candidates in a module are considered ($n$). For each candidate, a binary search is performed to a neighbouring module ($log(n)$), and for each doublet a binary search to the third module is performed ($log(n)$). Finally, the best hit within the threshold ($n$) is found. Therefore, its worst-case complexity is $O(m \cdot n^2 \cdot log(n)^2)$.

Track following occurs on all modules considered for extrapolation ($m$), where all forming tracks are considered ($n$). Similarly to seeding, the last doublet in each track is extrapolated and the best hit within the threshold is kept. Therefore, its worst-case complexity is $O(m \cdot n^2 \cdot log(n))$.

\subsection*{Tracklet filter}

The \emph{Tracklet filter} operates on the \emph{tracklet} container, which contains three-hit tracks for which no compatible hit was found in any track following step. It performs a least means square fit over the three hit coordinates of each track, and requires the fit be under a configurable threshold, and the three hits be not flagged, prior to accepting the tracklets as valid tracks. The worst-case complexity of the tracklet filter is $O(m \cdot n)$.
\section{Results}

% Results:
% * Search by triplet performs the tracking but it's embedded in the full VELO reconstruction sequence
%   The times it takes to do the three steps described here in the overall picture is as follows...
% * The following CPU / GPUs were considered (table with main features). This is the performance, these are some other figures like consumption, etc.
% * We are relatively close to the PCIe3 bandwidth cap (600 / 800 kHz)

The algorithms composing Search by triplet are not run in a separate application, but rather they are embedded as part of the VELO reconstruction sequence. An in-depth discussion of the additional algorithms involved in the sequence is out of the scope of this paper. Nevertheless, to provide an overall perspective of our work we present the VELO reconstruction sequence, which is composed of the following sequence of algorithms:

\begin{itemize}
\item Global event cut -- Rejects the 10\% most densely populated events.
\item Data transmission -- Copies VELO raw data from the host to the device.
\item VELO decoding -- Decodes the VELO raw data into hits.
\item VELO tracking -- Our Search by triplet implementation.
\item Data consolidation -- Packs resulting track data to improve memory locality of subsequent reconstruction algorithms.
% \item Optional validation -- Retrieves tracks from the device and validates them against the Monte Carlo truth.
\end{itemize}

The VELO reconstruction sequence is written in the Allen framework, which can be compiled for a variety of heterogeneous target devices. The resulting application can be run with a configurable number of events per bunch $n$ and threads $t$. Each thread instantiates an SIMT stream\footnote{When targeting a CPU architecture, the concept of stream is irrelevant and all algorithm invocations are run by the thread.}, and the thread-stream pair executes the requested sequence over the number of events. We run the presented experiments thousands of times to ensure reproducibility and mitigate processor warm-up effects.

\begin{figure}[hbt!]
\centering

\includegraphics[width=\textwidth]{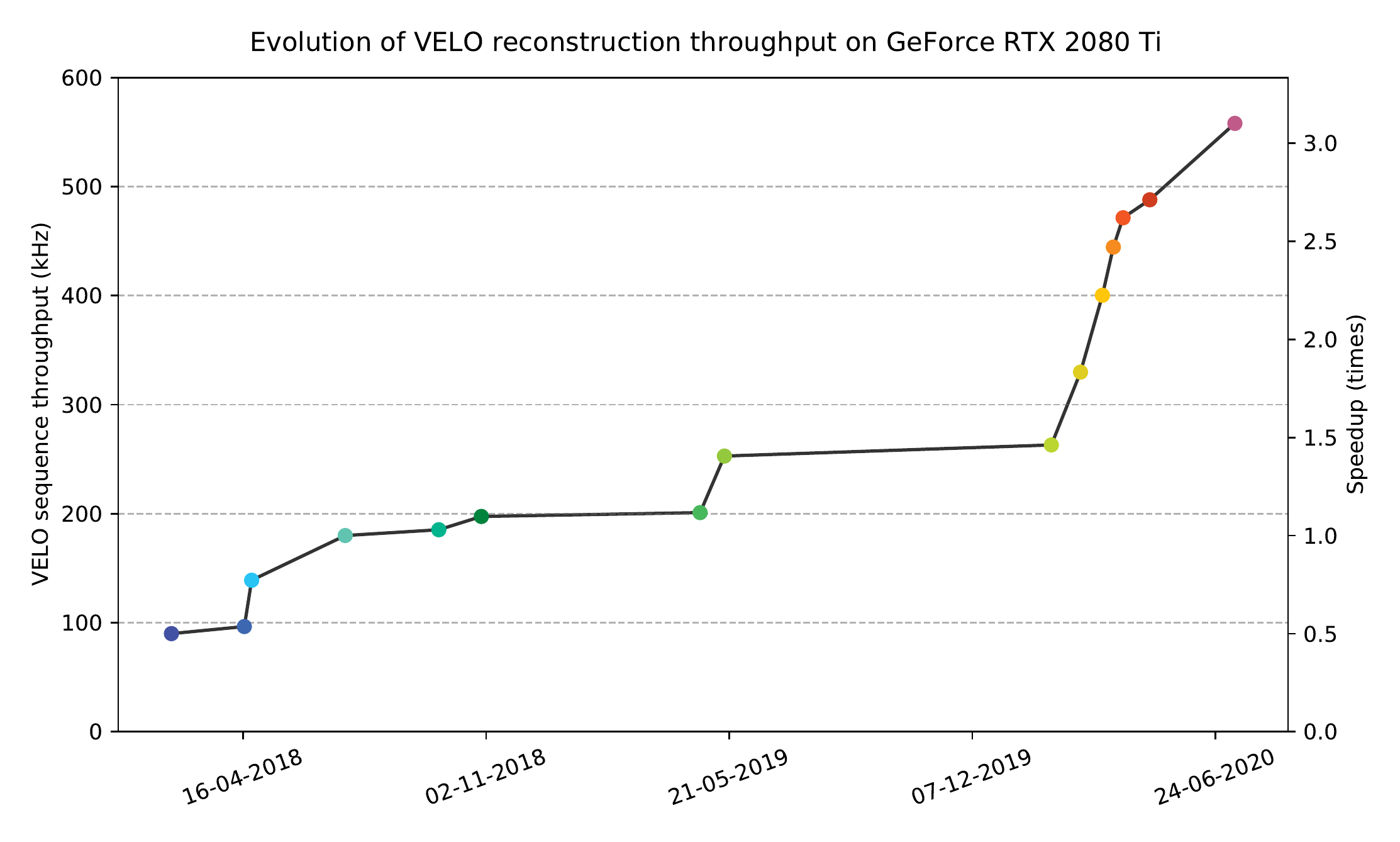}

\resizebox{\columnwidth}{!}{%
\begin{tabular}[!hbt]{|c|c|c|c|c|}
\hline
Date & Icon & Commit & Optimizations & Speedup \\\hline
16-02-2018 & 
\begin{tikzpicture}
\definecolor{colorpoint0}{HTML}{4251A3}
\draw (0,0) -- (0.6,0);
\fill[colorpoint0] (0.3,0) circle (0.1cm);
\end{tikzpicture} & \href{https://gitlab.cern.ch/lhcb/Allen/-/commit/1b30486e}{1b30486e} & Modern Search by triplet implementation. & $1.00\times$\\\hline
17-04-2018 & 
\begin{tikzpicture}
\definecolor{colorpoint1}{HTML}{3E67B1}
\draw (0,0) -- (0.6,0);
\fill[colorpoint1] (0.3,0) circle (0.1cm);
\end{tikzpicture} & \href{https://gitlab.cern.ch/lhcb/Allen/-/commit/6edc1109}{6edc1109} & Use shared memory in sorting algorithm. & $1.07\times$\\\hline
23-04-2018 & 
\begin{tikzpicture}
\definecolor{colorpoint2}{HTML}{27C4F4}
\draw (0,0) -- (0.6,0);
\fill[colorpoint2] (0.3,0) circle (0.1cm);
\end{tikzpicture} & \href{https://gitlab.cern.ch/lhcb/Allen/-/commit/b3218740}{b3218740} & Set statically maximum number of registers per kernel to 64. & $1.44\times$\\
& & & Store hit indices as 16-bit integers. & \\\hline
09-07-2018 & 
\begin{tikzpicture}
\definecolor{colorpoint3}{HTML}{60C4B1}
\draw (0,0) -- (0.6,0);
\fill[colorpoint3] (0.3,0) circle (0.1cm);
\end{tikzpicture} & \href{https://gitlab.cern.ch/lhcb/Allen/-/commit/6e74ee3f}{6e74ee3f} & Separate \emph{Fill candidates} and \emph{Tracklet filter} kernels. & $1.30\times$\\
& & & Reference implementation used in~\cite{CamporaPerez2019}. & \\\hline
24-09-2018 & 
\begin{tikzpicture}
\definecolor{colorpoint4}{HTML}{00B48D}
\draw (0,0) -- (0.6,0);
\fill[colorpoint4] (0.3,0) circle (0.1cm);
\end{tikzpicture} & \href{https://gitlab.cern.ch/lhcb/Allen/-/commit/8d337446}{8d337446} & Use a three-hit class for tracklets. & $1.03\times$\\\hline
29-10-2018 & 
\begin{tikzpicture}
\definecolor{colorpoint5}{HTML}{00833D}
\draw (0,0) -- (0.6,0);
\fill[colorpoint5] (0.3,0) circle (0.1cm);
\end{tikzpicture} & \href{https://gitlab.cern.ch/lhcb/Allen/-/commit/bd446eef}{bd446eef} & Cut 10\% busiest events. & $1.07\times$\\\hline
27-04-2019 & 
\begin{tikzpicture}
\definecolor{colorpoint6}{HTML}{48B85C}
\draw (0,0) -- (0.6,0);
\fill[colorpoint6] (0.3,0) circle (0.1cm);
\end{tikzpicture} & \href{https://gitlab.cern.ch/lhcb/Allen/-/commit/98b1cfff}{98b1cfff} & Use simplified scatter function. & $1.02\times$\\
& & & Offload prefix sum to CPU. & \\\hline
17-05-2019 & 
\begin{tikzpicture}
\definecolor{colorpoint7}{HTML}{95C93D}
\draw (0,0) -- (0.6,0);
\fill[colorpoint7] (0.3,0) circle (0.1cm);
\end{tikzpicture} & \href{https://gitlab.cern.ch/lhcb/Allen/-/commit/6e1f9b75}{6e1f9b75} & Remove static maximum number of registers per kernel. & $1.26\times$\\\hline
10-02-2020 & 
\begin{tikzpicture}
\definecolor{colorpoint8}{HTML}{BCD631}
\draw (0,0) -- (0.6,0);
\fill[colorpoint8] (0.3,0) circle (0.1cm);
\end{tikzpicture} & \href{https://gitlab.cern.ch/lhcb/Allen/-/commit/2869e656}{2869e656} & Ensure const-ness of input datatypes. & $1.04\times$\\\hline
05-03-2020 & 
\begin{tikzpicture}
\definecolor{colorpoint9}{HTML}{DFCE21}
\draw (0,0) -- (0.6,0);
\fill[colorpoint9] (0.3,0) circle (0.1cm);
\end{tikzpicture} & \href{https://gitlab.cern.ch/lhcb/Allen/-/commit/f7e85e0f}{f7e85e0f} & Use \texttt{fp-16} to store VELO hit coordinates. & $1.25\times$\\
& & & Use AOS for VELO hit coordinates. & \\\hline
23-03-2020 & 
\begin{tikzpicture}
\definecolor{colorpoint10}{HTML}{FFC60B}
\draw (0,0) -- (0.6,0);
\fill[colorpoint10] (0.3,0) circle (0.1cm);
\end{tikzpicture} & \href{https://gitlab.cern.ch/lhcb/Allen/-/commit/11853438}{11853438} & Move sources of each compilation unit to single files. & $1.21\times$\\\hline
01-04-2020 & 
\begin{tikzpicture}
\definecolor{colorpoint11}{HTML}{F68B1F}
\draw (0,0) -- (0.6,0);
\fill[colorpoint11] (0.3,0) circle (0.1cm);
\end{tikzpicture} & \href{https://gitlab.cern.ch/lhcb/Allen/-/commit/47d7348}{47d7348} & Adopt pendulum search for initial doublet. & $1.11\times$\\
& & & Remove \emph{Fill candidates} kernel. & \\\hline
09-04-2020 & 
\begin{tikzpicture}
\definecolor{colorpoint12}{HTML}{F15623}
\draw (0,0) -- (0.6,0);
\fill[colorpoint12] (0.3,0) circle (0.1cm);
\end{tikzpicture} & \href{https://gitlab.cern.ch/lhcb/Allen/-/commit/8038636}{8038636} & Use module pairs instead of single modules. & $1.06\times$\\
& & & Store hit phi in 16-bit integer. & \\\hline
01-05-2020 & 
\begin{tikzpicture}
\definecolor{colorpoint13}{HTML}{CF3D1E}
\draw (0,0) -- (0.6,0);
\fill[colorpoint13] (0.3,0) circle (0.1cm);
\end{tikzpicture} & \href{https://gitlab.cern.ch/lhcb/Allen/-/commit/f93df89e}{f93df89e} & Tune default number of threads to 64. & $1.03\times$\\\hline
10-07-2020 & 
\begin{tikzpicture}
\definecolor{colorpoint14}{HTML}{C05A89}
\draw (0,0) -- (0.6,0);
\fill[colorpoint14] (0.3,0) circle (0.1cm);
\end{tikzpicture} & \href{https://gitlab.cern.ch/lhcb/Allen/-/commit/fe529217}{fe529217} & Set as many active connections as streams. & $1.14\times$\\
& & & Use simplified atan2 function. & \\\hline
\end{tabular}
}

\caption{Evolution of VELO throughput versus time on a GeForce RTX 2080 Ti card.}
\label{fig:throughput-evolution}
\end{figure}

We have observed the performance evolution of the VELO reconstruction sequence over time. We have focused on the GPU~performance, which will perform the first stage of HLT in LHCb. Figure~\ref{fig:throughput-evolution} shows the evolution of the performance with each of the optimizations we have applied. The reported throughput numbers were obtained with the configuration $n = 1000$ and $t = 20$ in a GeForce~RTX~2080~Ti GPU. In all cases, the applications were compiled with the NVIDIA compiler $nvcc$. We report the throughput of Search by triplet since its modern inception in February~2018, and we align the speedup to the July~2018 version used for our previous publication~\cite{CamporaPerez2019}. For each optimization, the relevant branch of the code, a brief explanation of the optimization and its relative speedup are presented.

From July 2018 to July 2020, the performance of the entire VELO reconstruction sequence has improved by more than a factor~$3 \times$. Some optimizations have been discussed throughout section~\ref{sec:design}. One notable speedup has resulted from using array-of-structures (AOS) and moving to 16-bit floating point precision (\texttt{fp-16}) for storing and accessing VELO hit coordinates (05-03-2020, $1.25 \times$). AOS improved memory locality and coalescing since the three hit coordinates were always requested as a group. \texttt{fp-16} provides enough precision for the VELO coordinate range while decreasing memory pressure.

Three relevant sources of speedup did not involve code transformations. We first added the $nvcc$ option $maxrregcount$ to statically set the maximum number of registers of our kernels to 64, which improved performance (23-04-2018). However, the code evolved and we later disabled it, observing a speedup of $1.26 \times$ (17-05-2019). We think this is likely due to the separation of kernels on 09-07-2018, which allowed the compiler to set a better value to improve execution on each of them separately the second time.

Another such speedup resulted from adopting \emph{no separable compilation} (23-03-2020, $1.21 \times$). Separable compilation allows compilation units to use functions that are defined but not declared within their scope. Alongside obtaining a substantial speedup, our code became compatible with AMD~GPUs as well through use of the HIP compiler. Currently it would be possible to improve the performance of separable compilation codebases with Link Time Optimization (LTO), but the release of $nvcc$ at the time did not have this feature. Finally, setting as many active connections as streams (10-07-2020) improved the performance when setting configurations with more than eight active threads.

We have closely tracked the physics efficiency of our algorithm throughout its development cycle by means of continuous integration tests. The physics efficiency of our algorithm has either remained constant or improved with every iteration. In particular, the only cut introduced in our sequence (27-04-2019, $1.07 \times$) is a configurable cut that removes events with high detector occupancy and consequently low signal-to-background ratio, which is common practice in real-time, high-throughput, resource-limited physics reconstruction environments.

Table~\ref{tab:physics-efficiency} shows the physics performance indicators for a variety of particle categories, for both the reference implementation and the latest version of Search~by~triplet. The different particle categories refer to different decay processes. For the LHCb experiment of special interest are decays of Beauty particles indicated by \emph{From B} and also decays from so-called strange particles or into electrons. The physics performance indicators were introduced in Section~\ref{sec:introduction}.

\begin{table}[hbt!]
\begin{center}
\resizebox{\textwidth}{!}{
\begin{tabular}{|l|c|c|c|c||c|c|c|c|}
\hline
% Overall fake fraction   & 3.25 & & & \\\hline
                        & \multicolumn{4}{c||}{Reference (09-07-2018)} & \multicolumn{4}{c|}{Latest version (10-07-2020)} \\\hline
\multicolumn{1}{|c|}{Particle category} & Reco. & Clone & Hit & Hit & Reco. & Clone & Hit & Hit \\
                        & eff.  & fraction & purity & eff. & eff. & fraction & purity & eff. \\\hline
All                     & 95.36 & 0.60 & 99.07 & 97.58 & 98.52 & 2.14 & 99.30 & 96.45 \\\hline
% Long                    & 97.31 & 0.57 & 99.22 & 97.96 & 99.41 & 1.47 & 99.56 & 97.79 \\\hline
% Long, $p > 5$ GeV         & 98.51 & 0.51 & 99.23 & 98.24 & 99.65 & 1.18 & 99.66 & 98.41 \\\hline
Strange                 & 87.45 & 0.60 & 98.04 & 97.65 & 98.13 & 1.58 & 99.48 & 97.35 \\\hline
% Long strange, $p > 5$ GeV & 91.45 & 0.49 & 97.96 & 98.14 & 98.37 & 0.94 & 99.39 & 98.55 \\\hline
From B                  & 96.88 & 0.51 & 98.96 & 97.85 & 99.30 & 1.16 & 99.74 & 98.11 \\\hline
% Long from B, $p > 5$ GeV  & 98.42 & 0.43 & 98.96 & 98.02 & 99.56 & 1.00 & 99.77 & 98.43 \\\hline
Electrons               & 70.25 & 0.72 & 90.93 & 90.64 & 97.38 & 2.74 & 98.18 & 97.02 \\\hline
From B electrons        & 79.00 & 1.13 & 94.00 & 93.52 & 97.00 & 3.68 & 98.42 & 96.68 \\\hline
% Long from B electrons, p > 5 GeV & 77.97 & 1.45 & 94.50 & 93.93 \\\hline\\\hline
% Long from B electrons, p > 5 GeV & 97.44 & 3.73 & 98.63 & 97.42 \\\hline
Overall fake fraction   & \multicolumn{4}{c||}{3.25} & \multicolumn{4}{c|}{0.86} \\\hline
\end{tabular}
}
\end{center}
\caption{Efficiency numbers of VELO reconstruction sequence for reference implementation (left) and latest version (right). For each implementation, the reconstruction efficiency, clone fraction, hit purity and hit efficiency are shown as percentages for a variety of particle categories. The last row shows the overall fake fraction of the implementation as a percentage over the total number of reconstructed tracks.}
\label{tab:physics-efficiency}
\end{table}

Reconstruction efficiency has substantially improved across all categories. The tighter scatter criterium has reduced fake fraction at the cost of increasing clone fraction. The decay products of strange particles are especially challenging to reconstruct as they leave fewer hits in the VELO detector and do not necessarily point to the interaction region, as they are produced as secondary decay products from relatively long-lived particles. We have improved the reconstruction efficiency of decay products of strange particles by more than 10\%. We have also improved the efficiency for electron tracks by around 20\%, while increasing their hit efficiency and purity as well.

\begin{figure}[hbt!]
\centering
\includegraphics[width=\textwidth]{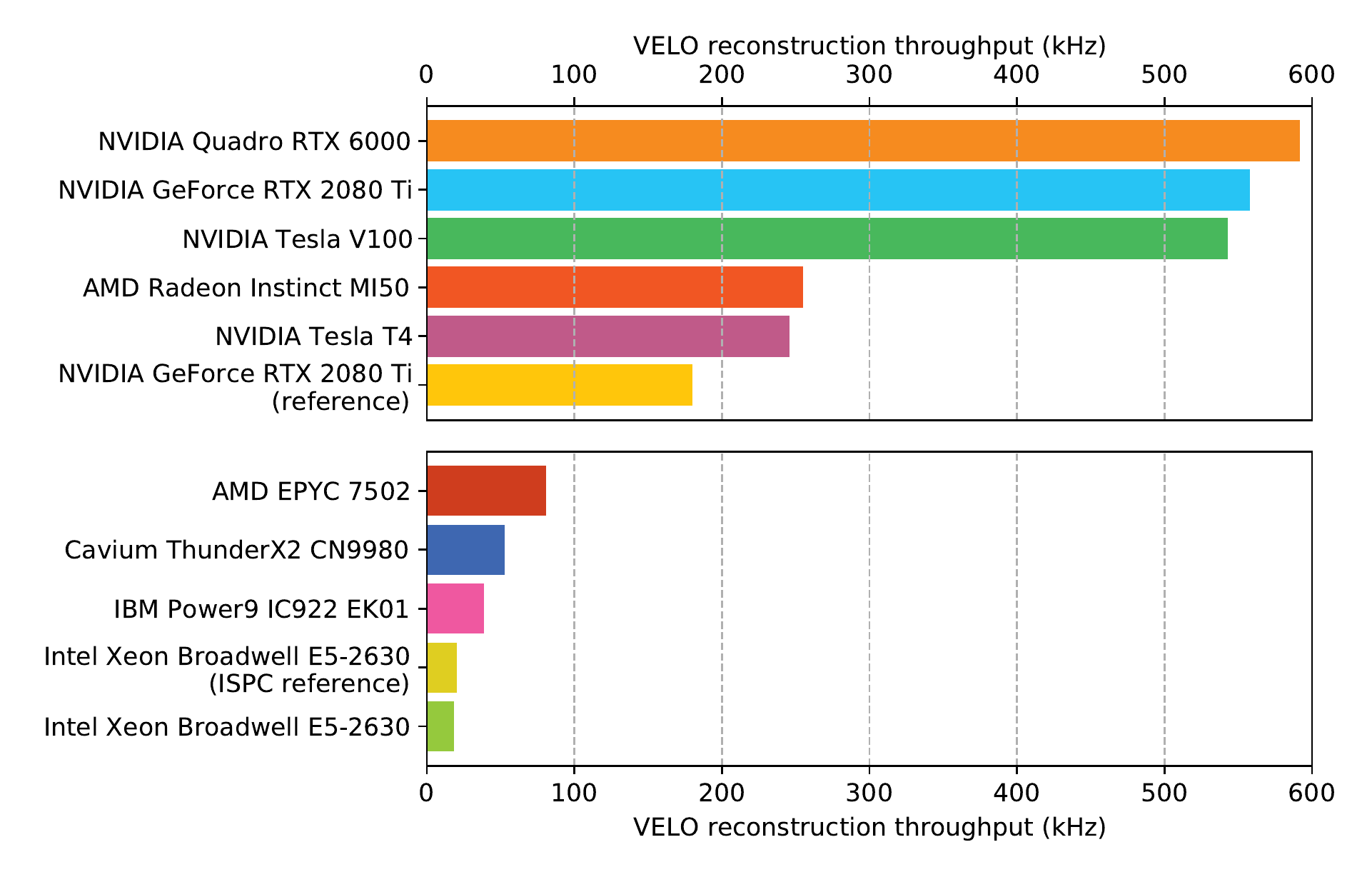}
\caption{Throughput of full VELO reconstruction sequence, including global event cut, data transmission, decoding, tracking and data consolidation. (Top) Throughput across GPUs. (Bottom) Throughput across single-socket CPUs. The reference performance from~\cite{CamporaPerez2019} is included for both CPU and GPU.}
\label{fig:velo-throughput}
\end{figure}

The entire sequence has been run over a variety of architectures, listed in table~\ref{tab:hw}. Our tests have run over four CPUs from different vendors and five GPUs. The throughput of the VELO reconstruction sequence across architectures is shown in Figure~\ref{fig:velo-throughput}. GPU architectures are depicted on the top, whereas CPU architectures are depicted on the bottom.

\begin{table}[hbt!]
\begin{center}
\resizebox{\textwidth}{!}{
\begin{tabular}{|l|c|c|c|c|c|c|c|}
\hline
Feature & Type & \# cores & Freq. (GHz) & Peak TFLOPS & MSRP (\$) & TDP (w) \\\hline
Intel Xeon Broadwell E5-2630 & CPU & 10 (20) & 3.1 & 0.352 & 667 & 85 \\\hline
IBM Power9 IC922 EK01 & CPU & 16 (64) & 4.0 & 0.512 & 3150 & 225 \\\hline
Cavium ThunderX2 CN9980 & CPU & 32 (128) & 2.5 & 0.56 & 1795 & 180 \\\hline
AMD EPYC 7502 & CPU & 32 (64) & 3.35 & 1.3 & 3618 & 180 \\\hline
AMD Radeon Instinct MI50 & GPU & 3840 & 1.746 & 13.41 & 8999 & 300 \\\hline
NVIDIA GeForce RTX 2080 Ti & GPU & 4352 & 1.545 & 13.34 & 1199 & 250 \\\hline
NVIDIA Quadro RTX 6000 & GPU & 4608 & 1.77 & 16.31 & 4000 & 250\\\hline
NVIDIA Tesla T4 & GPU & 2560 & 1.59 & 8.141 & 2295 & 70 \\\hline
NVIDIA Tesla V100 32GB & GPU & 5120 & 1.37 & 14.13 & 8999 & 250 \\\hline
\end{tabular}
}
\end{center}
\caption{Hardware used for our tests.}
\label{tab:hw}
\end{table}

The tests are run with a configuration of $n = 500$ and $t = 16$ for all GPUs. For CPUs, the configuration is $n = 200$ and $t$ has been set to the number of simultaneous multi-threads of each processor. We observe best performance on the Quadro~RTX~6000 card. If we take into account the price from table~\ref{tab:hw}, the best price performance is found with the GeForce~RTX~2080~Ti, which is a consumer-grade product. If we consider the rate per TDP, best performing device is the Tesla~T4 due to its low TDP of 75~watts.

The performance of our previous ISPC~\footnote{The Intel Implicit SPMD Program Compiler~\cite{SPMD2} is an SPMD compiler that exposes vector units in SIMD processors as programmable processors. ISPC programs are written in a custom extension to the C language.} implementation is also presented for reference. Our current implementation is 10\% slower than our previous ISPC implementation. However, the ISPC implementation was a separate codebase that would introduce maintainance overhead, especially considering the lack of C++ support of ISPC. In addition, ISPC only support x86-based architectures and ARM32. In contrast, we currently employ standard C++ and manual vectorization through the vectorization library UMESIMD~\cite{Karpinski2017}, which supports x86 SSE, AVX and AVX512 extensions, Altivec, ARM64 and a compatibility scalar backend. In the future, we intend to transition to the standard library simd~\cite{Kretz2015}, currently under development.

% TODO: Actually I don't know about the branching ratio.

Search by triplet constitutes 70\% of the sequence time on CPUs, but only 47\% of the GPU sequence time. This difference likely has to do with differences in the execution model, as the branching ratio affects CPUs and GPUs differently. We observe a difference in performance of an order of magnitude between the CPUs and GPUs under consideration. The specifications of the Radeon~Instinct~MI50 match those of the Tesla~V100 in terms of TFLOPS, however we observe a performance gap of $2.13 \times$ in favor of the NVIDIA card. We have used the HIP compiler to generate the application for AMD GPUs, which is currently in active development and we have observed performance improvements with the recent transition from \texttt{hipcc} to \texttt{hip-clang}. We will continue to track the performance in AMD devices with the evolution of the compiler.

The tested generation of NVIDIA cards connect through the standard PCIe3 $16\times$, which has a peak payload bandwidth of 15.76~GBps~FDx. Our current application uses 8~GBps host to device data transfer, so theoretically we are currently using 51\% of the available bandwidth in PCIe3 $16\times$. We obtain a throughput of 592~kHz on the Quadro~RTX~6000, and the peak throughput attainable with only data transmission is 940~kHz, therefore empirically we are using 63\% of the capacity of the link. While in the last two years we have improved the throughput of our algorithm by more than $3 \times$, it will not be possible to improve it further by more than $1.6 \times$ in the future under these conditions. The next generation of NVIDIA cards will support PCIe4, which should enable us to overcome this potential issue.

\begin{figure}[hbt!]
\centering
\includegraphics[width=0.9\textwidth]{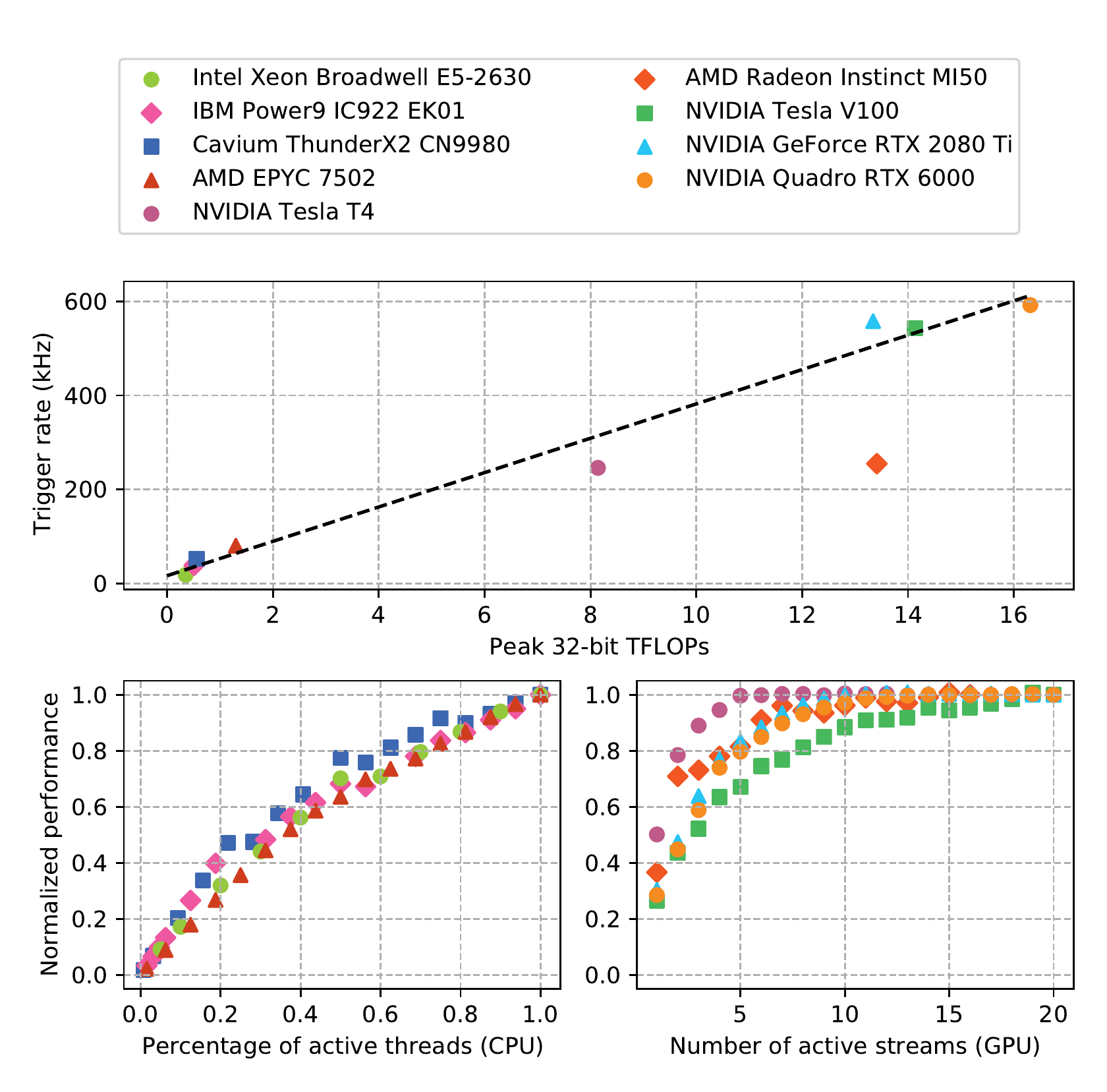}
\caption{(Top) VELO throughput versus peak 32-bit TFLOPs of processors under analysis. (Bottom left) Scalability of our program on CPUs under analysis. The x~axis refers to the number of active threads in the program with respect to the total number of SMTs of the processor. (Bottom right) Scalability of our program on GPUs with an increasing number of active streams.}
\label{fig:velo-multivalues}
\end{figure}

Figure~\ref{fig:velo-multivalues} shows the scalability of our application against the peak 32-bit FLOPS of the processors under consideration and against the number of threads. The top shows cross-architecture scalability with respect to TFLOPS of each system. Our software scales with the TFLOPS of the processors under consideration. The AMD~GPU appears as an outlier which indicates that more performance can be attained from further optimizing the code for that device.

The scalability of CPU processors is depicted in the bottom left. The run configurations were locked to $n = 200$, with a varying number of threads $t$ run in a single processor. For each CPU, the best performance was achieved by running a configuration with as many threads as simultaneous multi threads (SMTs). The AMD~EPYC and Intel processors show a steady performance increase with every additional core, and their scalability changes when they start using its SMTs. Both the Power and ARM processors exhibit four local peaks, which correspond with their 4-way SMT configuration.

Finally, the scalability of GPU processor is shown in the bottom right. For the GPU tests, each configuration was run with $n = 500$ and $t$ in the range~$[1, 20]$. The Tesla~T4 saturates with as few as 5~streams. The AMD card performance fluctuates and saturates with 14~streams. Both the GeForce and the Quadro cards scale very similarly, which could be due to both sharing the NVIDIA~Turing architecture. This is further supported by the different scaling behaviour of the Tesla~V100, of the previous Volta architecture. All GPUs saturate with a configuration of at least 14~streams. Our software scales to both CPU and GPU architectures.

\section{Conclusions}

We have presented Search by triplet, a fast algorithm for VELO reconstruction on parallel architectures. Our algorithm exploits various degrees of parallelism in LHCb VELO track reconstruction, and makes an efficient use of resources in heterogeneous architectures. The algorithm is written in a single codebase in C++, and we have developed architecture-specific optimizations for hot sections of the code.

% Our algorithm builds on the foundation of our previous paper

Our algorithm employs a local tracking technique to detect particle trajectories. The algorithm first sorts hits in each module by $\varphi$, generating an efficient 1D search structure. The local tracking method then consists in an iterative process with track seeding and track following steps. In track seeding, triplet seeds are sought as compatible non-flagged hits minimizing a fit function. Forming tracks are extended in track following, which flags hits of extended tracks. The flagging mechanism introduces a data dependency between seeding and following stages expressed as a barrier in SIMT architectures. Finally, remaining three-hit tracklets are filtered in a post-processing step. We expect our design to be generalizable to track reconstruction of other detectors, but this is subject to further research.

We have optimized our algorithm in the last two years iteratively, resulting in more than a factor~$3 \times$ speedup. We have described and quantified the speedup of each of the individual optimizations. We have discussed the inclusion of our algorithm as part of the VELO reconstruction sequence, and we have shown the improvement over time of the sequence. We have analyzed the performance of our algorithm over a variety of CPU and GPU architectures with respect to our previous implementations. We trade off a loss of 10\% in performance in our CPU implementation with the better maintainability of a single codebase and the portability across architectures of our new algorithm. We have tracked and improved the physics efficiency of our algorithm.

We have also discussed the price performance and throughput against TDP of the processors analyzed. We observe a peak throughput of 592~kHz on the Quadro~RTX~6000, the best price performance on the GeForce~RTX~2080~Ti, and the best throughput against TDP on the Tesla~T4. We observe it is possible to increase throughput by another roughly 40\% before PCIe3 bandwidth becomes a bottleneck.

We have shown that our algorithm scales with the FLOPS of the processors analyzed. Our algorithm also scales with the number of processors of the various processors under consideration. We have shown that a configuration with 14~streams saturates the performance in the GPUs under consideration.

Our algorithm constitutes the state-of-the-art in VELO reconstruction on SIMT architectures. The algorithm will be included in the HLT1 sequence of real-time software reconstruction in the LHCb data acquisition system, expected to restart data taking in late 2021. The throughput improvements presented gives the HLT1 margin to deal with the increase in data volume. The performance of the VELO reconstruction is crucial to the correct functioning of the LHCb experiment, we will continue to explore techniques to obtain better performance in current and upcoming hardware generations.
\section{Acknowledgements}

The authors would like to acknowledge the support of the LHCb collaboration throughout the development of the Search by triplet algorithm. We thank the LHCb Online team for the hardware support during our tests. We would also like to thank the LHCb computing, RTA and simulation teams for their support and for producing the simulated LHCb samples used to develop and benchmark our algorithm. We thank R.~Schwemmer for fruitful discussions about the performance of our algorithm. We thank R.~Aaij and T.~Suerink for their help in setting up tests in the Nikhef Power and AMD servers. We thank D.~vom~Bruch for providing the datasets used for tests. We thank X.~Valls~Pla for his help in setting up tests on the Cavium~ThunderX2 platform. We thank NVIDIA and especially A.~Hehn for the support and fruitful discussions.

\bibliographystyle{IEEEtran}
\bibliography{IEEEabrv,/home/dcampora/cernbox/phd/papers/mendeley_bibtex/library}

\end{document}